\documentclass{article}
\usepackage{setspace}
\usepackage[pdftex]{graphicx}
\usepackage{amsmath}
\usepackage{amssymb}
\usepackage{bbm}
\usepackage{graphicx}
\usepackage[backend=bibtex]{biblatex}
\begin{document}

\title{A pliable lasso for the Cox model}
\author{
  Wenfei Du\\
  Department of Statistics\\
  Stanford University\\
  \and
  Rob Tibshirani\\ 
  Departments of Biomedical Data Science\\
  and Statistics\\
  Stanford University\\
}
\maketitle

\begin{abstract}
We introduce a pliable lasso method for estimation of interaction effects in the Cox proportional hazards model framework.  The pliable lasso is a linear model  that includes interactions between covariates $X$ and a set of modifying variables $Z$ and assumes sparsity of the main effects and interaction effects.  The hierarchical penalty excludes interaction effects when the corresponding main effects are zero: this avoids overfitting and an explosion of model complexity.   We extend this method to the Cox model for survival data, incorporating modifiers that are either fixed or varying in time into the partial likelihood.  For example, this allows modeling of survival times that differ based on interactions of genes with age, gender, or other demographic information.  The optimization is done by blockwise coordinate descent on a second order approximation of the objective.
\end{abstract}

\section{Introduction}

We consider an approach to estimating high dimensional covariate and interaction effects in the survival analysis framework.  For example, a common case occurs with patient survival time data, which may include genetic information and demographics.  We might use a model that includes interactions between these two components.  The assumption is that the effect of gene expression may differ depending on variables like the age or gender of the individual.  The problem is that the set of genes and thus the set of possible interactions may be large.  We propose to use a pliable lasso penalty to incorporate these interactions in a hierarchical way.  This allows for interactions among the genes that have a main effect on survival.  Another example is if the relationship between genes and survival changes over time.  Using the pliable lasso, we can estimate possible gene interactions with time and time dependent variables as well.

The pliable lasso was introduced in the Gaussian setting in \cite{tibshirani2017}.  In this setting, suppose the data contains high dimensional covariates $x_i$, modifying variables $z_i$, and outcomes $y_i$.  The regression model including interactions between each $x_i$ and each $z_i$ is
\[ y_i = \beta_0 - z_i\theta_0 - \sum_{k=1}^px_{ik}(\beta_k + z_i\theta_k) + \epsilon_i \]
where $\epsilon_i$ is a noise term assumed to be Gaussian.  Since $x_i$ is high dimensional, regularized least squares regression can be used to estimate main effects $\beta_k$ and interaction effects $\theta_k$.  We can add a lasso penalty on $\beta_k$ or on $\beta_k$ and $\theta_k$.  The problem is that the first method ignores interactions and the second might not find main effects due to the large number of interaction terms.  Under the assumption that the interactions are hierarchical, \cite{tibshirani2017} instead proposed adding the pliable lasso penalty to get the objective
\[ M(\beta_0, \theta_0, \beta, \theta) = \frac{1}{2n}\sum_{i=1}^n(y_i - \beta_0 - z_i\theta_0 - \sum_{k=1}^px_{ik}(\beta_k + z_i\theta_k))^2 \]
\[ + \lambda((1 - \alpha)\sum_{k=1}^p(\left\|(\beta_k, \theta_k)\right\|_2 + \left\|\theta_k\right\|_2) + \alpha\sum_{k=1}^p\left\|\theta_k\right\|_1) \]
We optimize $M(\beta_0, \theta_0, \beta, \Theta)$ with respect to $\beta$ and $\theta$, where $\beta_0$ and $\theta_0$ are intercept and main effect terms for modifiers, $\beta$ is the vector of $\beta_k$ terms and $\theta$ is the vector of concatenated $\theta_k$ vectors.  This penalty term includes an overlapping group lasso penalty $\left\|(\beta_k, \theta_k)\right\|_2 + \left\|\theta_k\right\|_2$.  If $\theta_k$ is nonzero in the solution, $\beta_k$ is also nonzero, but $\theta_k$ can be zero when $\beta_k$ is nonzero.  This imposes a hierarchy on the main effects and corresponding interactions.

In the survival analysis setting, consider data with high dimensional covariates $x_i$, the time observed $y_i$, and whether the sample is censored $\delta_i$.  We are interested in estimating the effects of each variable on failure time.  The likelihood is 
\[ \prod_{i=1}^nf(y_i|x_i)^{\delta_i}S(y_i|x_i)^{1-\delta_i} \]
where $f$ is the density and $S$ the survival function.  In order to estimate the effects of interest, a parametrization can be used.  One often used model is Cox's proportional hazards model \cite{cox}, which defines the hazard function to be $h(t) = h_0(t)e^{x_i\beta}$.  Each individual has the same baseline hazard with a proportionality constant determined by their characteristics.  The limitation of this form is that the effect of a variable on survival is fixed across time.  The trend over time is determined only be $h_0(t)$ and is shared by all individuals.  The parameter of interest is $\beta$, which gives the effect of each covariate.  Under this model, we maximize the partial likelihood
\[ L(\beta) = \prod_{i=1}^m\frac{e^{x_{j(i)}\beta}}{\sum_{j\in R_i}e^{x_j\beta}} \]
where $j(i)$ is the index of the observation corresponding to time $i$ and $R_i$ is the risk set at time $i$.

One drawback of the model is that with high dimensional covariates the solution for $\beta$ may be degenerate.  Adding a lasso penalty is proposed in \cite{tibshirani1996} for linear regression and extended for Cox's model in \cite{tibshirani1997}.  A more general elastic net penalty
\[ \lambda P_\alpha(\beta) = \lambda(\alpha\sum_{k=1}^p|\beta_k| + (1/2)(1 - \alpha)\sum_{k=1}^p\beta_k^2) \]
can also be used.  Since the data is high dimensional, these problems are computationally difficult, and \cite{simon} proposes a fast coordinate descent algorithm.

In this paper, we consider a survival model including high dimensional covariates $x_i$ and their interactions with known modifying variables $z_i$.  We incorporates these interaction effects in a hierarchical way following \cite{tibshirani2017} using the pliable lasso.  We extend the blockwise coordinate descent algorithm and screening conditions in the paper to the proportional hazards model by approximating the partial log likelihood problem as a weighted least squares objective.  We also propose an application to the non proportional case, where time varying variables can be chosen as modifiers.  

\section{Cox's proportional hazards model}

We start with the Cox proportional hazard function $h(t) = h_0(t)e^{x_i\beta}$, in which the hazard function for the individuals are proportional to each other with covariates $x$ determining the level.  Supposing that the data does not have ties or weights, we include modifying variables $z$ and interactions $w_j = x_j \circ z$.  The Cox proportional hazard function with modifying variables is then
\[ h_i(t) = h_0(t)e^{z_i^t\theta_0 + x_i^t\beta + w_i^t\theta} \]
and the partial likelihood and log likelihood are
\[ L(\theta_0, \beta, \theta) = \prod_{i=1}^m\frac{e^{z_j(i)^t\theta_0 + \sum_{k=1}^px_{j(i)k}\beta_k + w_{j(i)k}^t\theta_k}}{\sum_{j\in R_i}e^{z_j^t\theta_0 + \sum_{k=1}^px_{jk}\beta_k + w_{jk}^t\theta_k}} \]
\[ l(\theta_0, \beta, \theta) = \sum_{i=1}^m z_j(i)^t\theta_0 + \sum_{k=1}^px_{j(i)k}\beta_k + w_{j(i)k}^t\theta_k - \log(\sum_{j\in R_i}e^{z_j^t\theta_0 + \sum_{k=1}^px_{jk}\beta_k + w_{jk}^t\theta_k}) \]
Since including interactions increases the dimension of parameters, the solution to the log likelihood maximization may not be well defined, particularly in the case of high dimensional covariates $x$.  We propose to use the pliable lasso penalty, and get the objective
\[ J(\theta_0, \beta, \theta) = \sum_{i=1}^m z_j(i)^t\theta_0 + \sum_{k=1}^px_{j(i)k}\beta_k + w_{j(i)k}^t\theta_k - \log(\sum_{j\in R_i}e^{z_j^t\theta_0 + \sum_{k=1}^px_{jk}\beta_k + w_{jk}^t\theta_k}) \]
\[ + \lambda((1 - \alpha)\sum_{k=1}^p(\left\|(\beta_k, \theta_k)\right\|_2 + \left\|\theta_k\right\|_2) + \alpha\sum_{k=1}^p\left\|\theta_k\right\|_1) \]
The advantage over the lasso penalty is that this allows hierarchical inclusion of interactions with $x_k$ depending on the inclusion of $x_k$ in the model.  The assumption is that variables that do not enter the model also do not have interactions but variables that do enter the model may or may not have interactions.

A coordinate descent algorithm for the least squares optimization with pliable lasso penalty is given in \cite{tibshirani2017}.  We follow the same optimization steps, but at each iteration of the algorithm, we approximate the Cox log likelihood using a quadratic expansion to obtain the weighted least squares problem.  Following \cite{simon}, we use a Taylor expansion at the current parameter values $(\tilde{\theta}_0, \tilde{\beta}, \tilde{\theta})$.  Setting $\tilde{\eta} = Z\tilde{\theta}_0 + X\tilde{\beta} + W\tilde{\theta}$, we have
\[ l(\theta_0, \beta, \theta) \approx l(\tilde{\theta}_0, \tilde{\beta}, \tilde{\theta}) \]
\[ + ((\theta_0, \beta, \theta) - (\tilde{\theta}_0, \tilde{\beta}, \tilde{\theta}))^t\dot{l}(\tilde{\theta}_0, \tilde{\beta}, \tilde{\theta}) + ((\theta_0, \beta, \theta) - (\tilde{\theta}_0, \tilde{\beta}, \tilde{\theta}))^t\ddot{l}(\tilde{\theta}_0, \tilde{\beta}, \tilde{\theta})((\theta_0, \beta, \theta) - (\tilde{\theta}_0, \tilde{\beta}, \tilde{\theta}))/2 \]
\[ \approx l(\tilde{\theta}_0, \tilde{\beta}, \tilde{\theta}) + (Z\theta_0 + X\beta + W\theta - \tilde{\eta})^tl'(\tilde{\eta}) + (Z\theta_0 + X\beta + W\theta - \tilde{\eta})^tl''(\tilde{\eta})(Z\theta_0 + X\beta + W\theta - \tilde{\eta})/2 \]
\[ = (Z\theta_0 + X\beta + W\theta - \tilde{\eta} + l''(\tilde{\eta})^{-1}l'(\tilde{\eta}))^tl''(\tilde{\eta})(Z\theta_0 + X\beta + W\theta - \tilde{\eta} + l''(\tilde{\eta})^{-1}l'(\tilde{\eta}))/2 + C(\tilde{\theta}_0, \tilde{\beta}, \tilde{\theta}) \]
Defining $\delta_j$ to be indicator of failure and $C_j$ to be the set of times where $j$ is at risk for $j$ a failure index, the derivatives of the partial log likelihood are 
\[ l'(\tilde{\eta})_j = \delta_j - \sum_{i\in C_j}\frac{e^{\eta_j}}{\sum_{k\in R_i}e^{\eta_k}} \]
\[ l''(\tilde{\eta})_{jj} = -\sum_{i\in C_j}\frac{(\sum_{k\in R_i}e^{\eta_k})e^{\eta_j}-e^{2\eta_j}}{(\sum_{k\in R_i}e^{\eta_k})^2} \]
The log likelihood in the objective may be approximated using only the diagonal elements of $l''$.  Adding the pliable lasso penalty to the approximation gives
\[ M(\theta_0, \beta, \theta) = \frac{1}{2n}\sum_{j=1}^n-l''(\tilde{\eta})_{jj}(\tilde{\eta}_j - l'(\tilde{\eta})_j/l''(\tilde{\eta})_{jj} - z_j^t\theta_0 - \sum_{k=1}^px_{jk}\beta_k - w_{jk}^t\theta_k)^2 \]
\[ + (1 - \alpha)\lambda\sum_{k=1}^p\left\|(\beta_k, \theta_k)\right\|_2 + \left\|\theta_k\right\|_2 + \alpha\lambda\sum_{k=1}^p\left\|\theta_k\right\|_1 \]
Optimizing $M(\theta_0, \beta, \theta)$ for $\beta_0$, $\beta$, and $\theta$ is a weighted least squares maximization problem with weights $-l''(\tilde{\eta})$
and adjusted residuals $r = (-\text{diag}(l''(\tilde{\eta})))(\tilde{\eta} - l'(\tilde{\eta})/\text{diag}(l''(\tilde{\eta})) - Z\theta_0 - X\beta - W\theta)$.  The problem may be solved using the Gaussian pliable lasso algorithm \cite{tibshirani2017}.  The advantage of this algorithm is that we evaluate conditions under which each $\beta_k$ and $\theta_k$ are zero based on the adjusted residuals before proceeding to the gradient descent step.  This saves computation for sparse models, where many of the effects are zero.  The details are given in the appendix.

In the general case, survival data may include ties for times of failure or weights for observations.  We use the Breslow approximation \cite{breslow}, which assumes the same denominator for each tied sample.  We weight each observation accordingly.  Suppose the weights are given by $\omega$, $D_i$ are the indices that fail for time $i$ and $d_i$ is the sum of the weights for these indices.  The partial likelihood and log likelihood are
\[ L(\beta, \theta) = \prod_{i=1}^m\frac{\prod_{j\in D_i}(\omega_je^{z_j^t\theta_0 + \sum_{k=1}^px_{jk}\beta_k + w_{jk}^t\theta_k})^{\omega_j}}{(\sum_{j\in R_i}\omega_j e^{z_j^t\theta_0 + \sum_{k=1}^px_{jk}\beta_k + w_{jk}^t\theta_k})^{d_i}} \]
\[ l(\beta, \theta) = \sum_{i=1}^m \sum_{j\in D_i}\omega_j(z_j^t\theta_0 + \sum_{k=1}^px_{jk}\beta_k + w_{jk}^t\theta_k) - d_i\log(\sum_{j\in R_i}w_je^{z_j^t\theta_0 + \sum_{k=1}^px_{jk}\beta_k + w_{jk}^t\theta_k}) \]
The derivatives are
\[ l'(\tilde{\eta})_j = \omega_j\delta_j - \sum_{i\in C_j}d_i\frac{\omega_je^{\eta_j}}{\sum_{k\in R_i}\omega_ke^{\eta_k}} \]
\[ l''(\tilde{\eta})_{jj} = -\sum_{i\in C_j}d_i\frac{(\sum_{k\in R_i}\omega_ke^{\eta_k})\omega_je^{\eta_j}-\omega_j^2e^{2\eta_j}}{(\sum_{k\in R_i}\omega_ke^{\eta_k})^2} \]
This is a generalization, and taking the weights to be one and failure times to be distinct returns the same log likelihood and derivatives as above.

For most applications, we are interested in determining a path of solutions corresponding to a sequence of $\lambda$ values.  Among these, cross validation can be used to select a model.  In this case, the maximum $\lambda$ of interest is when all of the coefficients are zero.  We approximate this $\lambda$ value by calculating the $\lambda$ at which all $\beta_k$ are zero, ignoring interactions.  This is the same calculation commonly used for lasso paths and is given by
\[ \lambda_{\max} = \max_k|X_k^t(-\text{diag}(l''(\tilde{\eta})))(\tilde{\eta} - l'(\tilde{\eta})/\text{diag}(l''(\tilde{\eta})) - Z\theta_0)/n| \leq (1 - \alpha)\lambda \]
To allow faster convergence at small $\lambda$ values, where the solution may not be sparse, warm starts are used.

Using the path of solutions, we would like to find the optimal $\lambda$ value.  Typically, $k$-fold cross validation is used, where validation on each of the $k$ folds is based on the log likelihood.  However, for the Cox model, since the partial log likelihood depends on the risk set sample, the value calculated using the validation fold may have few observations leading to a high variance estimate.  For the $i$th fold, we instead use the goodness of fit estimate from \cite{simon} given by
\[ l(\hat{\theta}_0, \hat{\beta}, \hat{\theta}) - l_{-i}(\hat{\theta}_0, \hat{\beta}, \hat{\theta}) \]
where $l_{-i}$ is the log likelihood on the rest of the folds not including $i$ and $\hat{\theta}_0$, $\hat{\beta}$, and $\hat{\theta}$ are the estimated coefficient values for each $\lambda$.  These estimates are then averaged over each of the folds and the minimizing $\lambda$ is chosen as optimal.

\section{Extension to the non-proportional hazards setting}

The assumption that the hazard functions for the individuals are proportional may be unrealistic.  The effect of a subset of the modifiers $z$ or covariates $x$ may be different across time, so that the hazard functions for differing values are not proportional.  A common generalization that accounts for interaction effects is to stratify based on the subset of non proportional variables, assuming it is known.  For individual $i$ in strata $g$, the hazard function is
\[ h_i(t) = h^g_0(t)e^{z_i^t\theta_0 + x_i^t\beta + w_i^t\theta} \]
where $z$ and $x$ now refers to the remaining covariates.  This allows the same effects $\theta_0$, $\beta$, and $\theta$ to be estimated across strata but a different baseline hazard $h^g_0$ for each.  To optimize, the stratified partial likelihood can be written with separate products for samples in each strata $g$.  Including a pliable lasso penalty, it is possible to generalize the estimation strategy above to this case.

We note that the advantage of stratification is that it allows for an interactive effect between time and the specified variables.  This suggests that pliable lasso may be used to select these variables if we treat time as a modifier in our model.  We consider the case when the subset of covariates with non proportional hazard functions are unknown.  We assume that the baseline hazard function depends on covariates $x$ through interactions between $x$ and some collection of functions $G(t)$.  The hazard function is
\[ h_i(t) = (h_0(t)e^{x_iG(t))^t\theta'_k})e^{z_i^t\theta_0 + x_i^t\beta + w_i^t\theta}.\]
We can approximate the interaction with time by taking $G(t)$ to be an evaluated spline bases.  Then, using a pliable lasso penalty with extended parameters $(\theta, \theta')$ instead of $\theta$, we are able to also estimate $\theta'$ and determine the non proportional covariates.  Combining this approach with the above stratification gives a more general model reflecting both known non proportional modifiers and unknown covariates.  In general, instead of a spline bases, the $G(t)$ function can be any time dependent modifiers of interest.  The estimation approach for this extension is given in the appendix.

To simplify the  computation,  we instead propose to use a logistic model to approximate the likelihood\footnote{This approximation appears to be known to the statistical community, but was new to us. We thank Terry Therneau for  very helpful discussions  about this issue.} .  For each failure time $i$, we create a covariate matrix by concatenating $x_{j(i)}$ with $x_j$ for each $j$ in the risk set.  We form the outcome vector by concatenating $y = 1$ indicating failure for observation $j(i)$ and $y = 0$ for the other observations.  The matrix for the possibly time dependent modifying variables is formed by concatenating the values of $x_{j(i)}G(i)$ with $x_jG(i)$ for each $j$ in the risk set.  Then, we stack the problems for each $i$ and use the binomial pliable lasso algorithm with separate intercepts for each time.  

To justify this procedure, we show that the logistic model log likelihood can approximate the Cox log likelihood.  The Cox log likelihood for observation $i$ is
\[ \eta_i - \log(\sum_{j \in R_i}e^{\eta_j}) \]
The logistic model log likelihood contribution for observation $i$ is 
\[ \beta_0 + \eta_i - \sum_{j \in R_i}\log(1 + e^{\beta_0}e^{\eta_j}) \]
Then, assuming $e^{\eta_j}/\sum_{j \in R_i}e^{\eta_j} \approx 0$, $\beta_0 = -\log(\sum_{j \in R_i}e^{\eta_j})$ solves the gradient condition and $-\sum_{j \in R_i}\log(1 + e^{\beta_0}e^{\eta_j}) \approx 0$.  This reduces to the Cox log likelihood contribution above.  The main assumption of this approximation is that the relative risk is near zero.  This holds if the risk set at each failure time is large, which happens when the data set is large and there are few failures. 

\section{Simulations}

We generate $x\sim N(0,1)$ and $z\sim \text{Bern}(0.5)$ with sample size $n = 100$, number of x variables $p = 10$, and number of z variables $nz = 4, 20$.  We use a constant baseline hazard $h_0(t)$, so that 
\[ h_i(t) = e^{z_i^t\theta_0 + x_i^t\beta + w_i^t\theta} \]
Define the linear term in the hazard as
\[ \lambda = x_1 - x_2 + x_3 + x_4 + x_1z_1 - x_1z_2 - 2x_2z_3 + 2x_2z_4 \]
For this specification, failure times follow an exponential distribution.  The model is
\[ y \sim e^{-\lambda}\text{Exp}(1) \]
and censoring times are generated by
\[ c \sim \text{Exp}(1). \]
All nonzero interactions are on variables with main effects.  Since the interactions are hierarchical, pliable lasso has an advantage.  The results for an alternative model that is not hierarchical are shown in the appendix.

The methods we compare are pliable lasso, lasso with no interactions, and lasso with all interactions using standardized covariates.  The lasso versions are optimized with {\tt glmnet} \cite{simon}.  The pliable lasso uses a fixed $\alpha = 0.5$, but we obtain similar results using a range of $\alpha$ values.  The cross validated model is chosen for each using the $\lambda$ value that minimizes the goodness of fit criteria listed above.  The $\lambda$ value using the one standard deviation rule is also tested, but the performance in terms of test negative log likelihood is worse for all methods.

The tables below include the negative log likelihood values evaluated using a test set of size $n = 1000$.  The value at the true coefficients are also shown.  In terms of this measure, pliable lasso performs better than both lasso methods for both values of $nz$.  For small $nz$, pliable lasso and the full lasso give comparable results.  The total number of variables including interactions is $50$ in this simulation, which is small enough for the full lasso to perform well.  However, for large $nz$, pliable lasso outperforms the full lasso, because the total number of variables is $210$, but performs similarly to the main effect lasso.  Both methods that include interactions do not select interaction effects well.  The hierarchical pliable lasso penalty enforces that the interactions are nonzero only when the main effect is nonzero.  Even though the model does relatively well on selecting main effects, the false positives and negatives increase the errors in selecting interaction effects.  On the other hand, without the hierarchical structure of the penalty, the lasso with interactions misses more main effects.  To have better selection of interaction effects, we may use a smaller model than the cross validated one.

\begin{table}[ht]
\centering
\begin{tabular}{rrrrrr}
  \hline
 & negative ll & fp ($\beta$) & fn ($\beta$) & fp ($\theta$) & fn ($\theta$) \\
  \hline
plasso & 2850.318 & 2.900 & 0.000 & 8.750 & 2.400 \\ 
lasso (main) & 3021.358 & 1.550 & 0.200 &  &  \\ 
lasso (full) & 2874.967 & 0.750 & 0.900 & 7.900 & 0.950 \\ 
true & 2702.726 &  &  &  &  \\ 
   \hline
\end{tabular}
\caption{\em Comparison of plasso with lasso: n = 100, p = 10, nz = 4}
\end{table}

\begin{table}[ht]
\centering
\begin{tabular}{rrrrrr}
  \hline
 & negative ll & fp ($\beta$) & fn ($\beta$) & fp ($\theta$) & fn ($\theta$) \\
  \hline
plasso & 3120.071 & 0.800 & 0.750 & 13.750 & 3.350 \\ 
lasso (main) & 3140.337 & 1.100 & 0.400 &  &  \\ 
lasso (full) & 3209.842 & 0.050 & 3.050 & 6.500 & 2.850 \\ 
true & 2647.001 &  &  &  &  \\ 
   \hline
\end{tabular}
\caption{\em Comparison of plasso with lasso: n = 100, p = 10, nz = 20}
\end{table}

We next consider a baseline hazard that includes interactions on covariates.  Using the interaction model in the previous section, this introduces time dependent modifiers.  We generate $x\sim \text{Unif}(0,1)$ and do not include any fixed modifiers for simplicity.  The number of x variables is $p = 10$, but we increase the sample size to $n = 500$.  This is chosen to compensate for the increased complexity of having modifiers that change with time.  The hazard we use is
\[ h_i(t) = e^{(x_it)^t\theta'}e^{x_i^t\beta} \]
Define the linear terms $x_i^t\beta$ as
\[ \lambda = x_1 - x_2 + x_3 + x_4 \]
and the time dependent linear terms $(x_it)^t\theta'$ as
\[ k = 5x_1 + 5x_2 \]
Failure times in this case follow the same distribution as the logarithm of an exponential variable.  The model is
\[ y \sim \log(\text{Exp}(1)e^\lambda k + 1) / k \]
and censoring times are generated by
\[ c \sim \text{Exp}(1). \]
All time dependency is on variables with main effects, giving pliable lasso an advantage.  The results for an alternative model that is not hierarchical are shown in a later section.

We compare the pliable lasso to the  lasso with no interactions and regular Cox estimation with time dependent variables.  For the pliable lasso, we use five linear spline terms, standardize all covariates including the splines, and sample risks sets of size five.  We present the results for pliable lasso using a fixed $\alpha = 0$.  With larger $\alpha$, pliable lasso outperforms the other methods in terms of log likelihood and main effect determination but does not find time effects.  Cox estimation with a linear time interaction using the {\tt coxph} R library  is included as a comparison since {\tt glmnet} does not allow possible time dependencies.  Cross validation with the goodness of fit criteria is used to choose the model for lasso and pliable lasso.

The tables below include the same measures as above, where the negative log likelihood is calculated with a test set of size $1000$.  For pliable lasso, the false positives and false negatives for $\theta$ are calculated based on selection of any of the spline terms corresponding to a given covariate interaction.  The log likelihood values are comparable, but the pliable lasso outperforms the lasso in selecting main effects.  It also does well in selecting interaction effects, but since Cox estimation does not set effects to zero, we cannot compare.

\begin{table}[ht]
\centering
\begin{tabular}{rrrrrr}
  \hline
 & negative ll & fp ($\beta$) & fn ($\beta$) & fp ($\theta$) & fn ($\theta$) \\ 
  \hline
plasso & 4902.511 & 1.550 & 0.100 & 2.000 & 0.500 \\ 
  lasso (main) & 4892.897 & 2.900 & 0.150 &  &  \\ 
  cox (full) & 4900.188 &  &  &  &  \\ 
  true & 4869.943 &  &  &  &  \\ 
   \hline
\end{tabular}
\caption{\em Comparison: n = 500, p = 10, nz = 5}
\end{table}

\section{NKI data example}

We test the Cox pliable lasso on the breast cancer NKI data from \cite{vantveer}\cite{vandevijver}.  The data contains survival information and gene expression levels of 24,481 genes for 319 individuals.  We consider distant metastasis free survival as the outcome and threshold on the variance of the genes to obtain 134 total covariates.  The lasso and pliable lasso estimators are compared, where the pliable lasso includes a linear time interaction.  The results shown below are from left out test data.

We use the logistic approximation to estimate the survival model.  The below plot shows the AUC for the test predictions from logistic lasso and logistic pliable lasso.  We see that the pliable lasso outperforms the lasso in terms of classifying survival events.

\begin{figure}[h!]
\begin{center}
\includegraphics[width=0.8\textwidth]{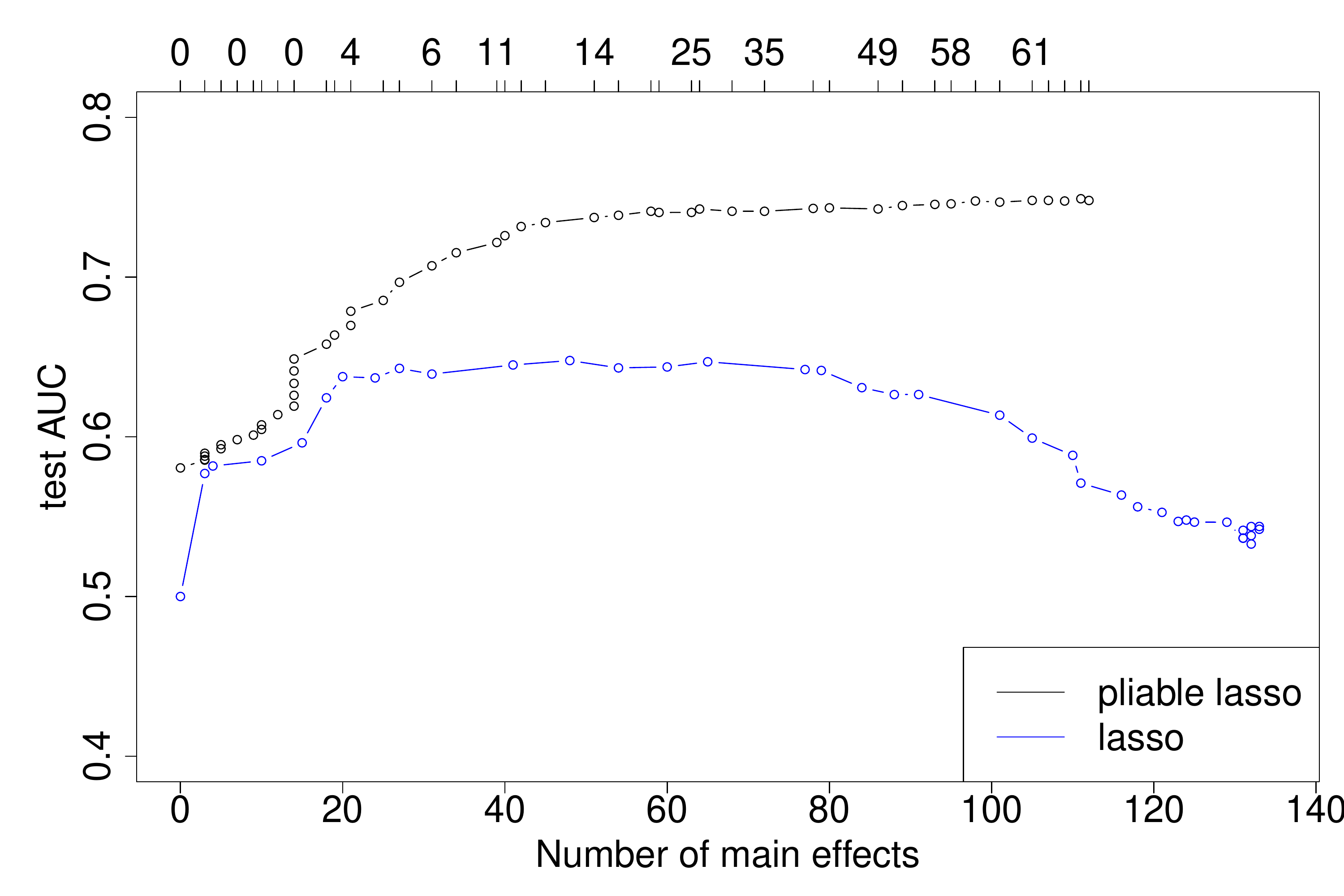}
\end{center}
\caption{\em Test set AUC for NKI data}
\end{figure}

The plot below shows that the Cox partial log likelihood for the test data is also larger for the pliable lasso estimates than for the lasso ones either using the approximate logistic model or the Cox model.  Adding the time varying component improves on estimates of the survival model as well as the classification.

\begin{figure}[h!]
\begin{center}
\includegraphics[width=0.8\textwidth]{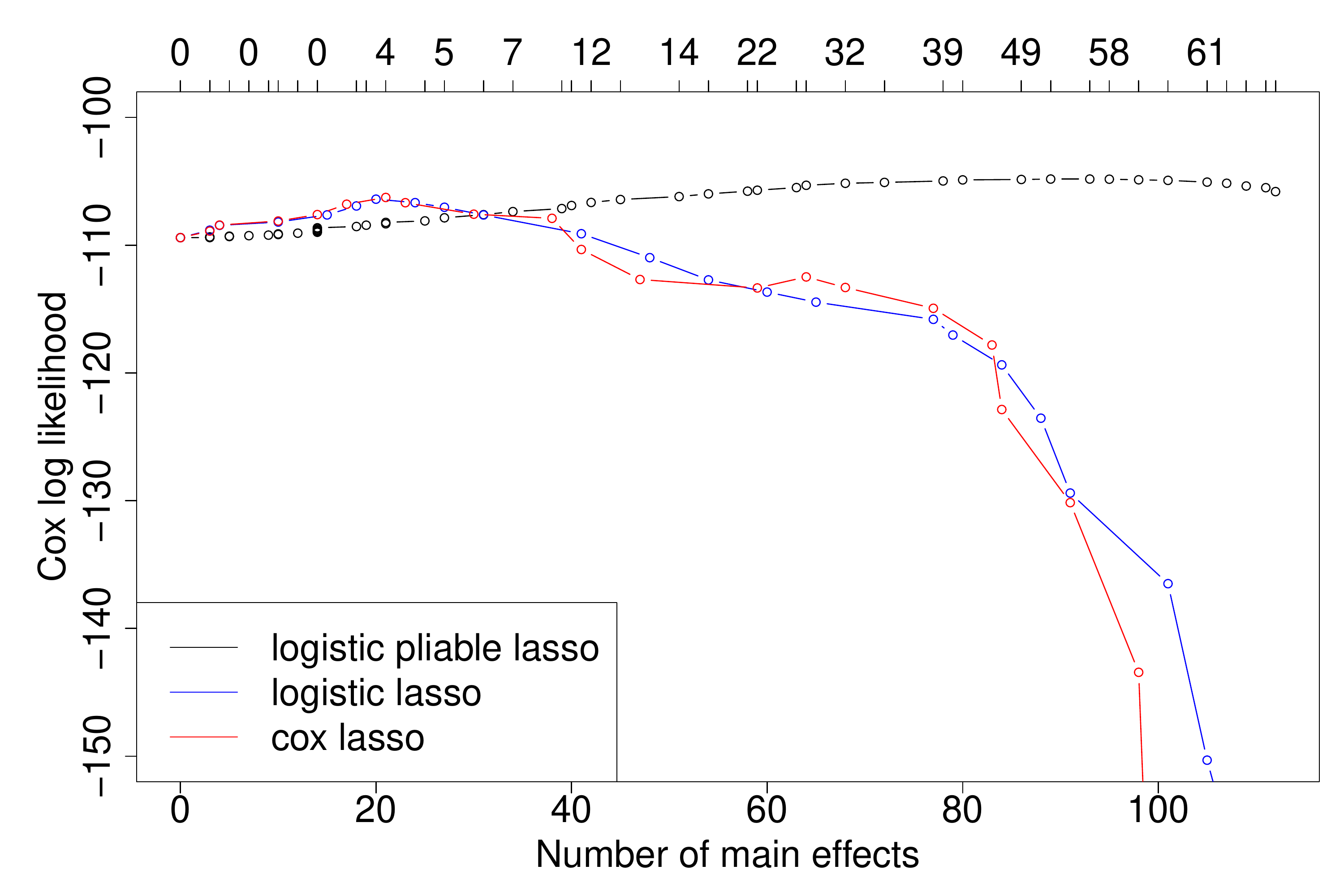}
\end{center}
\caption{\em Test set Cox log likelihood for NKI data}
\end{figure}

\section{Conclusion}

We have extended the  pliable lasso approach to Cox's model for survival data.  This allows for estimation of main effects for covariates and interaction effects between covariates and modifiers that may be fixed in time or time varying.  The estimation uses a blockwise coordinate descent algorithm that optimizes a weighted least squares problem with the penalty term at each step.  To simplify computations, particularly for the time varying case, we propose to use a logistic model to approximate the partial likelihood.  The implementation will be made available in R as an option in the plasso package.

\printbibliography

\section*{Appendix: Computational details for the non proportional hazards model}

The problem with the time dependent form of the hazard function is that the likelihood terms for the observations in the risk set in the denominator change depending on the time $i$.  For simplicity, we write $x$ and $w_j = x_j \circ B(t)$ and omit the $z$ variable terms.  The logic for calculating main effects and interactions on $z$ follows the same pattern.  We write the hazard function as
\[ h(t) = h_0(t)e^{x_j\beta + w_{jt}\theta} \]
and the partial likelihood and log likelihood as
\[ L(\beta, \theta) = \prod_{i=1}^m\frac{e^{\sum_{k=1}^px_{j(i)k}\beta_k + w_{j(i)ik}\theta_k}}{\sum_{j\in R_i}e^{\sum_{k=1}^px_{jk}\beta_k + w_{jik}\theta_k}} \]
\[ l(\beta, \theta) = \sum_{i=1}^m\sum_{k=1}^px_{jk}\beta_k + w_{jik}\theta_k - \log(\sum_{j\in R_i}e^{\sum_{k=1}^px_{jk}\beta_k + w_{jik}\theta_k}) \]
Due to the difference in the calculation of the risk set depending on $i$, the likelihood now depends on $\tilde{\eta}_j^0 = \sum_{k=1}^px_{jk}\tilde{\beta}_k + w_{jik}\tilde{\theta}_k$ and $\tilde{\eta}_{ji'}^1 = \sum_{k=1}^px_{jk}\tilde{\beta}_k + w_{ji'k}\tilde{\theta}_k$ for $i>i'$.  Labeling the corresponding matrices, we write $\tilde{\eta}^0 = X^0\beta + W^0\tilde{\theta}$ and $\tilde{\eta}^1 = X^1\beta + W^1\tilde{\theta}$.  The Taylor expansion is similar to the above but the derivatives are in terms of $\tilde{\eta} = (\tilde{\eta}^0, \tilde{\eta}^1)$.  We have
\[ l(\beta, \theta) \approx l(\tilde{\beta}, \tilde{\theta}) + ((\beta, \theta) - (\tilde{\beta}, \tilde{\theta}))^t(\dot{l}(\tilde{\beta}, \tilde{\theta}) )+ ((\beta, \theta) - (\tilde{\beta}, \tilde{\theta}))^t\ddot{l}(\tilde{\beta}, \tilde{\theta})((\beta, \theta) - (\tilde{\beta}, \tilde{\theta}))/2 \]
\[ \approx l(\tilde{\beta}, \tilde{\theta}) + ((X^0\beta + W^0\theta, X^1\beta + W^1\theta) - \tilde{\eta})^tl'_{\tilde{\eta}}(\tilde{\eta}) + ((X^0\beta + W^0\theta, X^1\beta + W^1\theta) - \tilde{\eta})^tl''_{\tilde{\eta}}(\tilde{\eta}) \]
\[ \times ((X^0\beta + W^0\theta, X^1\beta + W^1\theta) - \tilde{\eta})/2 \]
\[ = ((X^0\beta + W^0\theta, X^1\beta + W^1\theta) - \tilde{\eta} + l''(\tilde{\eta})^{-1}l'(\tilde{\eta}))^tl''(\tilde{\eta}) \]
\[ \times ((X^0\beta + W^0\theta, X^1\beta + W^1\theta) - \tilde{\eta} + l''(\tilde{\eta})^{-1}l'(\tilde{\eta}))/2 + C(\tilde{\beta}, \tilde{\theta}) \]
The derivatives are
\[ l'_{\tilde{\eta}^0}(\tilde{\eta})_j = \delta_j - \frac{e^{\eta_j + \eta_{ji}}}{\sum_{k\in R_i}e^{\eta_k + \eta_{ki}}} \]
\[ l''_{\tilde{\eta}^0}(\tilde{\eta})_{jj} = -\frac{(\sum_{k\in R_i}e^{\eta_k + \eta_{ki}})e^{\eta_j + \eta_{ji}}-e^{2(\eta_j + \eta_{ji})}}{(\sum_{k\in R_i}e^{\eta_k + \eta_{ki}})^2} \]
\[ l'_{\tilde{\eta}^1}(\tilde{\eta})_{ji'} = -\frac{e^{\eta_j + \eta_{ji'}}}{\sum_{k\in R_{i'}}e^{\eta_k + \eta_{ki}}} \]
\[ l''_{\tilde{\eta}^1}(\tilde{\eta})_{ji'ji'} = -\frac{(\sum_{k\in R_i'}e^{\eta_k + \eta_{ki'}})e^{\eta_j + \eta_{ji'}}-e^{2(\eta_j + \eta_{ji'})}}{(\sum_{k\in R_i'}e^{\eta_k + \eta_{ki'}})^2} \]
We again only using the diagonal elements of $l''$ in the Taylor approximation.  Adding the pliable lasso penalty gives
\[ M(\beta, \theta) = \frac{1}{2n}\sum_{j=1}^n-l''_{\tilde{\eta}^0}(\tilde{\eta})_{jj}(\tilde{\eta}^0_j - l'_{\tilde{\eta}^0}(\tilde{\eta})_j/l''_{\tilde{\eta}^0}(\tilde{\eta})_{jj}^0 - \sum_{k=1}^px_{jk}\beta_k - w_{jik}\theta_k)^2 \]
\[ + \frac{1}{2n}\sum_{ji'}-l''_{\tilde{\eta}^1}(\tilde{\eta})_{ji'ji'}(\tilde{\eta}_{ji'}^1 - l'_{\tilde{\eta}^1}(\tilde{\eta})_{ji'}/l''_{\tilde{\eta}^1}(\tilde{\eta})_{ji'ji'} - x_{jk}\beta - w_{ji'k}\theta_k)^2 \]
\[ + (1 - \alpha)\lambda\sum_{k=1}^p\left\|(\beta_k, \theta_k)\right\|_2 + \left\|\theta_k\right\|_2 + \alpha\lambda\sum_{k=1}^p\left\|\theta_k\right\|_1 \]
This is another weighted least squares plus pliable lasso penalty problem that may be solved using the Gaussian pliable lasso algorithm.  The weights are $(-l''_{\tilde{\eta}^0}(\tilde{\eta}), -l''_{\tilde{\eta}^1}(\tilde{\eta}))$ and the adjusted residuals are $(r^0, r^1) = ((-\text{diag}(l''_{\tilde{\eta}^0}(\tilde{\eta})))(\tilde{\eta}^0 - l'_{\tilde{\eta}^0}(\tilde{\eta})/\text{diag}(l''_{\tilde{\eta}^0}(\tilde{\eta})) - X^0\beta - W^0\theta), (-\text{diag}(l''_{\tilde{\eta}^1}(\tilde{\eta})))(\tilde{\eta}_j^1 - l'_{\tilde{\eta}^1}(\tilde{\eta})/\text{diag}(l''_{\tilde{\eta}^1}(\tilde{\eta})) - X^1\beta - W^1\theta))$.

Since we now calculate separate denominators for each time, the computation time is increased by an order of $m$.  To reduce it, we sample points in the denominator when computing the log likelihood and its derivatives.  This sample risk set procedure is proposed by \cite{liddell} and asymptotically justified by \cite{goldstein}.  In practice, we use the logistic approximation described above, which is much faster and gives similar results.
   
\section*{Optimization details for proportional and non proportional hazards model}

Following \cite{tibshirani2017}, we solve the weighted least squares problem by checking zero conditions for each group corresponding to $x_j$ and its interactions.  For the proportional hazard case, the derivatives of the objective with the pliable lasso penalty are
\[ \frac{dM(\theta_0, \beta, \theta)}{d\theta_0} = \frac{1}{n}\sum_{j=1}^n(-z_j)(-l''(\tilde{\eta})_{jj})(\tilde{\eta}_j - l'(\tilde{\eta})_j/l''(\tilde{\eta})_{jj} - z_j^t\theta_0 - \sum_{k=1}^px_{jk}\beta_k - w_{jk}^t\theta_k) = 0 \]
\[ \hat{\theta}_0 = (\sum_{j=1}^nz_j(-l''(\tilde{\eta})_{jj})z_j)^{-1}(\sum_{j=1}^nz_j(-l''(\tilde{\eta})_{jj})(\tilde{\eta}_j - l'(\tilde{\eta})_j/l''(\tilde{\eta})_{jj} - \sum_{k=1}^px_{jk}\beta_k - w_{jk}^t\theta_k)) \]
\[ \frac{dM(\theta_0, \beta, \theta)}{d\beta_k} = \frac{1}{n}\sum_{j=1}^n(-x_{jk})(-l''(\tilde{\eta})_{jj})(\tilde{\eta}_j - l'(\tilde{\eta})_j/l''(\tilde{\eta})_{jj} - z_j^t\theta_0 - \sum_{k=1}^px_{jk}\beta_k - w_{jk}^t\theta_k) \]
\[ + (1-\alpha)\lambda u \]
\[ = -X_k^tr/n + (1 - \alpha)\lambda u = 0 \]
\[ \frac{dM(\theta_0, \beta, \theta)}{d\theta_k} = \frac{1}{n}\sum_{j=1}^n(-w_{jk})(-l''(\tilde{\eta})_{jj})(\tilde{\eta}_j - l'(\tilde{\eta})_j/l''(\tilde{\eta})_{jj} - z_j^t\theta_0 - \sum_{k=1}^px_{jk}\beta_k - w_{jk}^t\theta_k) \]
\[ + (1-\alpha)\lambda(u_2 + u_3) + \alpha\lambda v \]
\[ = -W_k^tr/n + (1 - \alpha)\lambda(u_2 + u_3) + \alpha\lambda v = 0 \]
\[ (\beta_k, \theta_k) \neq 0, u = \frac{\beta_k}{\left\|(\beta_k, \theta_k)\right\|_2}, u_2 = \frac{\theta_k}{\left\|(\beta_k, \theta_k)\right\|_2} \]
\[ (\beta_k, \theta_k) = 0, \left\|(u, u_2)\right\|_2 \leq 1 \]
\[ \theta_k \neq 0, u_3 = \frac{\theta_k}{\left\|\theta_k\right\|_2} \]
\[ \theta_k = 0, \left\|u_3\right\|_2 \leq 1 \]
\[ v = \text{sign}(\theta_j) \]
The conditions for $\hat{\beta}_k = 0, \hat{\theta}_k = 0$ are 
\[ |X_k^tr_{(-k)}/n| \leq (1 - \alpha)\lambda \]
\[ \left\|S(W_k^tr_{(-k)}/n, \alpha\lambda)\right\|_2 \leq 2(1 - \alpha)\lambda \]
and the conditions for $\hat{\beta}_k \neq 0, \hat{\theta}_k = 0$ are
\[ \hat{\beta}_k = (n/\sum_{j=1}^n(-l''(\tilde{\eta})_{jj})(x_{jk})^2)S(X_k^tr_{(-k)}/n, (1 - \alpha)\lambda) \]
\[ \left\|S(W_k^t(r_{(-k)} - \text{diag}(l''(\tilde{\eta}))X_k\hat{\beta}_k)/n, \alpha\lambda)\right\|_2 \leq (1 - \alpha)\lambda \]
When $\hat{\beta}_k \neq 0, \hat{\theta}_k \neq 0$, taking $\gamma_k = (\beta_k, \theta_k)$, we optimize
\[ M(\gamma_k) = -l(\gamma_0) - (\gamma_k - \gamma_{0k})^t\dot{l}(\gamma_{0k}) + \frac{1}{2t}\left\|\gamma_k - \gamma_{0k}\right\|_2^2 + (1 - \alpha)\lambda\sum_{k=1}^p\left\|\gamma_k\right\|_2 + \left\|\theta_k\right\|_2 + \alpha\lambda\sum_{k=1}^p\left\|\theta_k\right\|_1 \]
following the gradient algorithm in \cite{tibshirani2017}.

For the non proportional hazard case, the derivatives are
\[ \frac{dM(\beta, \theta)}{d\beta_k} = \frac{1}{n}\sum_{j=1}^n(-x_{jk})(-l''_{\tilde{\eta}^0}(\tilde{\eta})_{jj})(\tilde{\eta}_j^0 - l'_{\tilde{\eta}^0}(\tilde{\eta})_j/l''_{\tilde{\eta}^0}(\tilde{\eta})_{jj} - \sum_{k=1}^px_{jk}\beta_k - w_{jik}^t\theta_k) \]
\[ + \frac{1}{n}\sum_{ji'}(-x_{jk})(-l''_{\tilde{\eta}^1}(\tilde{\eta})_{ji'ji'})(\tilde{\eta}_{ji'}^1 - l'_{\tilde{\eta}^1}(\tilde{\eta})_{ji'}/l''_{\tilde{\eta}^1}(\tilde{\eta})_{ji'ji'} - \sum_{k=1}^px_{jk}\beta_k - w_{ji'k}^t\theta_k) \]
\[ + (1-\alpha)\lambda u \]
\[ = -X_k^{0t}r^0/n - X_k^{1t}r^1/n + (1 - \alpha)\lambda u = 0 \]
\[ \frac{dM(\beta, \theta)}{d\theta_k} = \frac{1}{n}\sum_{j=1}^n(-w_{jik})(-l''_{\tilde{\eta}^0}(\tilde{\eta})_{jj})(\tilde{\eta}_j^0 - l'_{\tilde{\eta}^0}(\tilde{\eta})_j/l''_{\tilde{\eta}^0}(\tilde{\eta})_{jj} - \sum_{k=1}^px_{jk}\beta_k - w_{jik}^t\theta_k) \]
\[ + \frac{1}{n}\sum_{ji'}(-w_{ji'k})(-l''_{\tilde{\eta}^1}(\tilde{\eta})_{ji'ji'})(\tilde{\eta}_{ji'}^1 - l'_{\tilde{\eta}^1}(\tilde{\eta})_{ji'}/l''_{\tilde{\eta}^1}(\tilde{\eta})_{ji'ji'} - \sum_{k=1}^px_{jk}\beta_k - w_{ji'k}^t\theta_k) \]
\[ + (1-\alpha)\lambda(u_2 + u_3) + \alpha\lambda v \]
\[ = -W_k^{0t}r/n - W_k^{1t}r/n + (1 - \alpha)\lambda(u_2 + u_3) + \alpha\lambda v = 0 \]
\[ (\beta_k, \theta_k) \neq 0, u = \frac{\beta_k}{\left\|(\beta_k, \theta_k)\right\|_2}, u_2 = \frac{\theta_k}{\left\|(\beta_k, \theta_k)\right\|_2} \]
\[ (\beta_k, \theta_k) = 0, \left\|(u, u_2)\right\|_2 \leq 1 \]
\[ \theta_k \neq 0, u_3 = \frac{\theta_k}{\left\|\theta_k\right\|_2} \]
\[ \theta_k = 0, \left\|u_3\right\|_2 \leq 1 \]
\[ v = \text{sign}(\theta_j) \]
The zero condition checking and gradient optimizing steps are similar to the above.

\section*{Additional simulations}

We show results for the pliable lasso using a model that is not hierarchical.  Define the exponential parameter as
\[ \lambda = x_1 - x_2 + x_3 + x_4 + x_5z_1 - x_6z_2 - 2x_7z_3 + 2x_8z_4 \]
The rest of the specification is the same as in the previous model.  The tables below show similar results with increased number of false positives for $\beta$ and for $\theta$ for pliable lasso.  The full lasso outperforms pliable lasso in the case of small $nz$.  However, for large $nz$, pliable lasso again does slightly better than either lasso method.  However, comparing to the negative log likelihood value at the true model, all of the methods perform badly in this case.  

\begin{table}[ht]
\centering
\begin{tabular}{rrrrrr}
  \hline
 & negative ll & fp ($\beta$) & fn ($\beta$) & fp ($\theta$) & fn ($\theta$) \\
  \hline
plasso & 2804.377 & 5.000 & 0.000 & 9.100 & 2.400 \\ 
lasso (main) & 2956.584 & 4.350 & 0.150 &  &  \\ 
lasso (full) & 2722.543 & 1.150 & 0.000 & 10.050 & 0.050 \\ 
true & 2627.815 &  &  &  &  \\ 
   \hline
\end{tabular}
\caption{\em Comparison:$ n = 100, p = 10, nz = 4$}
\end{table}
   
\begin{table}[ht]
\centering
\begin{tabular}{rrrrrr}
  \hline
 & negative ll & fp ($\beta$) & fn ($\beta$) & fp ($\theta$) & fn ($\theta$) \\
     \hline
plasso & 3227.921 & 2.100 & 1.800 & 20.750 & 3.650 \\ 
lasso (main) & 3238.321 & 2.750 & 0.650 &  &  \\ 
lasso (full) & 3263.990 & 0.050 & 3.400 & 8.350 & 2.450 \\ 
true & 2639.533 &  &  &  &  \\ 
   \hline
\end{tabular}
\caption{\em Comparison: $n = 100, p = 10, nz = 20$}
\end{table}

We also test a time dependent model that is not hierarchical.  Define the shape parameter as 
\[ k = 5x_5 + 5x_6 \]
with the same specification as the previous model.  The main difference in the results is that the number of false positives for $\beta$ and for $\theta$ have increased for pliable lasso.  The full Cox model estimation outperforms pliable lasso in terms of negative log likelihood value, but including the time effects is still able to perform better than the model without them.

\begin{table}[ht]
\centering
\begin{tabular}{rrrrrr}
 & negative ll & fp ($\beta$) & fn ($\beta$) & fp ($\theta$) & fn ($\theta$) \\ 
  \hline
plasso & 4979.746 & 4.200 & 0.000 & 4.750 & 0.000 \\ 
lasso (main) & 4990.797 & 4.300 & 0.000 &  &  \\ 
cox (full) & 4965.987 &  &  &  &  \\ 
true & 4944.261 &  &  &  &  \\ 
   \hline
\end{tabular}
\caption{\em Comparison: $n = 500, p = 10, nz = 5$}
\end{table}

This provides evidence that the pliable lasso is robust to interactions that are not hierarchical.  The only case when a full lasso may be preferred is when the number of modifiers is small.  With a large number, the pliable lasso performs similarly to or better than the other methods in terms of negative log likelihood and determination of main effects.

\end{document}